\documentclass[11pt,a4paper]{article}

\usepackage[T1]{fontenc}
\usepackage[utf8]{inputenc}
\usepackage[english]{babel}
\usepackage[numbers,sort&compress]{natbib}
\usepackage{amsmath,amssymb,bm}
\usepackage{graphicx}
\usepackage{booktabs}
\usepackage{physics}
\usepackage{geometry}
\usepackage{setspace}
\usepackage{caption}
\usepackage{subcaption}
\usepackage{hyperref}
\usepackage[nameinlink,noabbrev]{cleveref}
\usepackage{authblk}

\DeclareCaptionLabelSeparator{verticalLine}{ | }
\hypersetup{
  colorlinks=true,
  linkcolor=blue,
  citecolor=blue,
  urlcolor=blue,
  pdftitle={Your Manuscript Title},
  pdfauthor={Author One, Author Two}
}
\title{\textbf{Vibrational high-harmonics and period-doubling bifurcation probed by time-resolved electron diffraction}}

\author[1]{Alexander Schröder}
\author[1]{Kai Nettersheim}
\author[1,2,3]{Ferdinand Evers}
\author[1,2,3,*]{Sascha Schäfer}

\affil[1]{Department of Physics, University of Regensburg, Regensburg, Germany}
\affil[2]{Regensburg Center for Ultrafast Nanoscopy (RUN), Regensburg, Germany}
\affil[3]{Center for Chiral Electronics, Halle/Berlin/Regensburg, Germany}
\affil[*]{Correspondence: \texttt{sascha.schaefer@ur.de}}

\date{} 

\begin{document}

\maketitle

\begin{abstract}
Nanoscale mechanical oscillators exhibit a plethora of nonlinear phenomena with promising applications for the sensing and clocking of processes down to atomic length scales. Oscillator dynamics are typically probed by electrical or optical means, providing only limited access to the spatial profile of the oscillator motion. Here, we introduce event-based convergent beam electron diffraction for the spatio-temporal mapping of nanoscale mechanical resonators in ultrafast transmission electron microscopy. Employing an optically driven silicon membrane resonator at various driving strengths, we gain access to nonlinear processes with increasing complexity, ranging from a simple Duffing behavior to nonlinear multimode coupling and period-doubling bifurcations. The time-resolved diffraction probing approach supports a spatial resolution down to a few nanometers and a temporal resolution of 5 ns and provides quantitative information on the local membrane bending. Because the diffraction signal responds to local displacement gradients, which become more pronounced as resonators shrink, this approach offers a route toward probing nonlinear nanomechanics at the atomic scale. 
\end{abstract}


\section*{Introduction} 
Micro- and nanoelectromechanical systems (MEMS/NEMS) are central to a wide range of applications, ranging from precision sensing \cite{Zhang.2013,Steeneken.2025} to resonant timing devices \cite{Uranga.2015} and signal processing \cite{Guerra.2010,Romero.2024}. In most cases, their performance is determined by the mechanical properties of the underlying nanostructures, in particular their resonance width, mode couplings and nonlinear dynamics. Understanding and controlling these vibrational behaviors is essential for optimizing stability, sensitivity, and energy efficiency. 

From a physical viewpoint, the nonlinear structural dynamics in nanomechanical resonators give rise to a variety of phenomena, including amplitude-dependent frequency shifts, bistability, modal coupling, and symmetry-breaking bifurcations up to period doubling and chaos \cite{Chowdhury.2017,Davidovikj.2017,Huber.2020,Keskekler.2021,L.Gammaitoni.1989,Karabalin.2009}. These effects are not only of fundamental interest but also offer practical opportunities for improved signal conversion, noise squeezing, and parametric amplification \cite{Suh.2010,Bachtold.2022}. Most readout schemes infer the structural motion in nanomechanical resonators indirectly from electrical or optical signals, leaving the underlying structural evolution, mode shapes, local strain fields, and inter-mode interactions only partially resolved. In both applied and fundamental cases, probing the resonator motion with a spatial resolution far below the optical diffraction limit would be highly beneficial.

Scanning electron microscopy in transmission and in backscattering geometries has been successfully employed for visualizing and controlling the motion of carbon nanotubes and cantilevers \cite{Buks.2001,Nigues.2015,Yasuda.2016,Tsioutsios.2017,Liu.2021,Cretu.2022} but still presents challenges with respect to a precise vibrational amplitude calibration and disentangling of electron-beam-induced effects from intrinsic system dynamics. Transmission electron microscopy (TEM) excels in the achievable spatial resolution and the quantitative extraction of structural distortions down to atomic length scale. Recent work has combined conventional TEM with the temporal resolution of femtosecond laser systems and allowed for the probing of ultrafast dynamics by femtosecond electron pulses -- a technique commonly referred to as ultrafast TEM (UTEM) \cite{Flannigan.2012,Feist.2017,Vanacore.2019,Kurman.2021,Cao.2021,Kim.2023,Harder.2023,Auad.2023,Meng.2024,Bucher.2024,Borrelli.2024,Gaida.2024,Gao.2025,Liu.2025,Schroder.2025}. With respect to structural dynamics, various UTEM approaches were employed for the mapping of acoustic phonon propagation and phase transitions by parallel and convergent beam electron diffraction, as well as bright or dark-field imaging \cite{Kwon.2008,Yurtsever.2011,Feist.2018,Zhang.2019,Kim.2020,Danz.2021,Bach.2022,Nakamura.2022,Xing.2025}. Following a pump-probe approach, in each of these cases, a single optical pulse initiates a structural response which decays before the next excitation pulse arrives. Thereby, large amplitude motion and pronounced intrinsic structural nonlinearities are typically not observed.

Here, we introduce a local electron probing methodology for resonantly excited, nonlinear structural dynamics in free-standing crystalline nanostructures with nanosecond temporal and nanometer spatial resolution utilizing an event-based probing approach. For high-Q silicon membrane resonators, we map the membrane's structural response to optical pulse trains in space and time. We are thus able to follow the distortions of the membrane evolving from a regime of purely harmonic excitation, through bistable Duffing behavior into nonlinear multimode coupling and self-induced parametrically driven period doubling. Besides the detailed mapping of fundamental nonlinear physical processes, the large structural distortion available in resonant pumping opens a novel driving mechanism in ultrafast transmission electron microscopy.  

\section*{Results}
\subsection*{Experimental Setup}
\begin{figure}[!htbp]
    \centering
    \includegraphics[width = 0.8\textwidth]{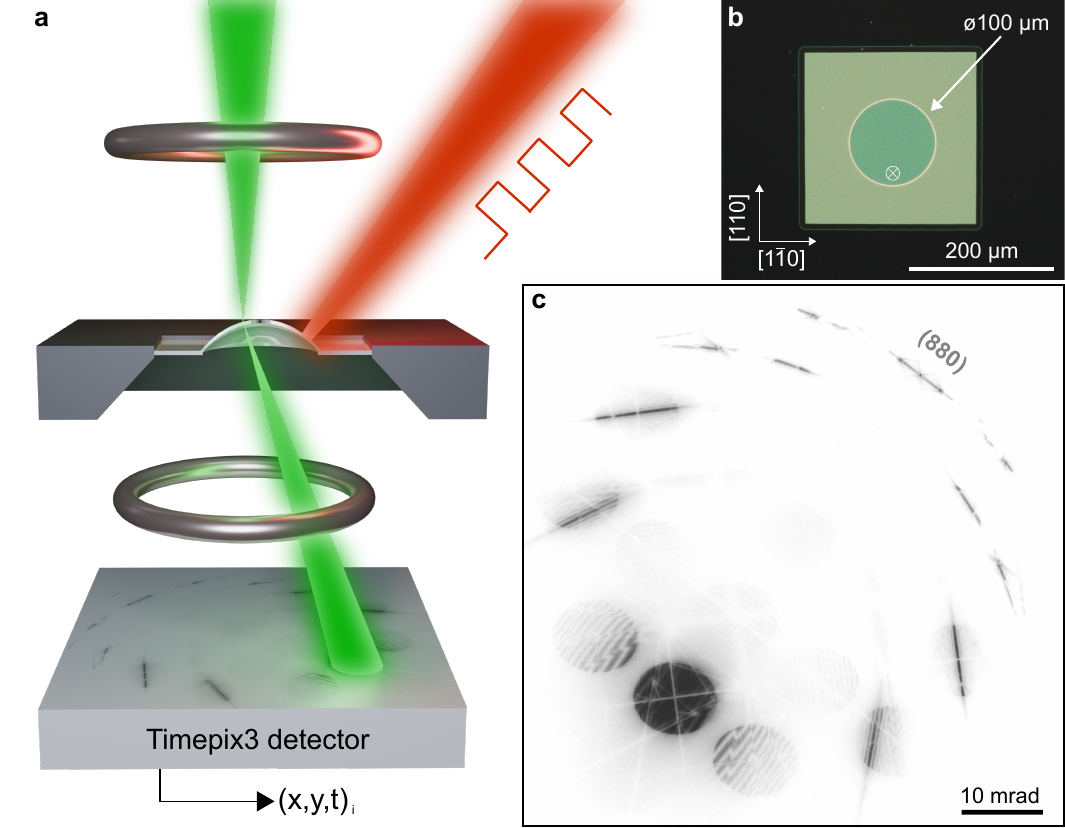}
    \caption{ \textbf{Experimental scheme and sample geometry}. \textbf{a} A convergent electron beam probes the local structural response in a free-standing, single-crystalline silicon membrane, optically driven by an amplitude modulated beam with a tunable modulation frequency in the MHz range. Electron arrival times at individual detector pixels are time-stamped with nanosecond precision and phase-sorted with respect to the optical modulation to reconstruct the time-resolved diffraction pattern. \textbf{b} Optical micrograph of the  membrane window suspended across a silicon frame (dark regions). The circular central region (dark green contrast) corresponds to the free-standing silicon membrane. In the outer rectangular region (light green contrast) an additional silicon-oxynitride overlayer is deposited, resulting in a uniform tensile stress state of the central membrane region. \textbf{c} Convergent beam electron diffraction pattern of the silicon membrane tilted by about 1.5 degree around the $[1\overline{1}0]$ direction.} \label{fig:1}
\end{figure}

To investigate the structural response of nanomechanical resonators with high spatial and temporal resolution, we combine convergent-beam electron diffraction (CBED) in a transmission electron microscope with nanosecond time-resolved electron detection (Fig. \ref{fig:1}a). The mechanical resonator is driven by an optical pulse train (785 nm wavelength) with tunable repetition rate, $1/T$, matched to a resonance frequency of the system. 
For probing the induced time-dependent local strain in the resonator, a continuous electron beam is focused on a selected position on the membrane (typical electron focal spot size:  20 nm) and electron diffraction patterns are recorded. The arrival time of each electron on the detector is tagged with a precision of about  5 ns \cite{Schroder.2024}, so that temporal changes in the diffraction pattern can be reconstructed with respect to the phase of the optical pulse train.

To showcase the capabilities of this approach for probing nonlinear structural dynamics, we investigate the structural response of a suspended, single-crystalline (100)-oriented silicon membrane (200-nm thickness). Lateral clamping by a silicon-oxynitride overlayer induces a symmetric tensile stress of about 200 MPa \cite{Shchepetov.2013,Masteghin.2021} in the central circular region of the membrane (100 µm in diameter) which is not covered by silicon oxynitride (Fig.~\ref{fig:1}b). In Fig. \ref{fig:1}c, a CBED pattern is shown recorded at the approximate position marked in Fig. \ref{fig:1}b and with the electron beam incident along the [001] zone axis. Each disk-shaped feature in the CBED pattern originates from the diffraction off a set of lattice planes corresponding to a reciprocal lattice spot with Bragg indices (hkl) \cite{Zuo.2017}. The diameter of each disk is given by the electron beam convergence angle, chosen as 10 mrad (full-angle) in our case. The intensity distribution within the disks reflects the angular dependent diffraction efficiency and thus maps a generalized rocking curve \cite{Zuo.2017} with the position of individual diffraction features sensitive to the orientation of the electron beam direction with respect to the local crystal orientation (see Supplemental Video 1 for a CBED tilt series). Therefore, a local bending of the silicon membrane after optical excitation results in a shift in the diffraction efficiency distribution with respect to the diffraction disk. For diffraction under larger angles and for tilted samples (Fig.~\ref{fig:1}c), the diffraction efficiency is localized along a line-like feature in each disk which shows additional intensity modulations along the line due to multiple scattering \cite{Zuo.2017}, as, for example, visible for the Bragg line (880) in Fig.~\ref{fig:1}c. Due to the ease of analysis, we focus in the following on the optically induced angular movement of the (880) Bragg line.

\subsection*{Duffing response}\label{sec3}

\begin{figure}[!htbp]
    \centering
    \includegraphics[width = 1\textwidth]{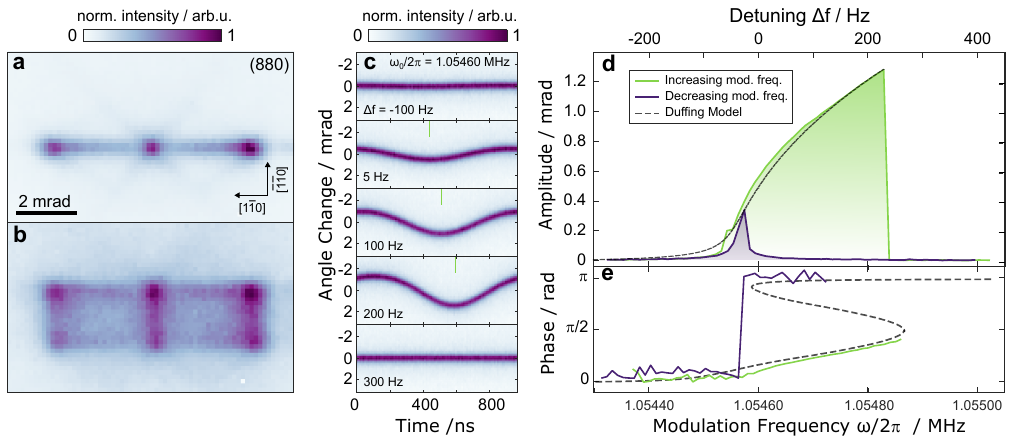}
    \caption{\textbf{Duffing-like response at low excitation power.} \textbf{a,b} Close-up of the temporally-averaged (880) diffraction line under optical excitation at two modulation frequencies (\textbf{a}: 1.05450 MHz, off resonance; \textbf{b}: 1.05470 MHz, on resonance). \textbf{c} Evolution of (880) Bragg line profiles over one drive period for different modulation frequencies. The profile maps are obtained by integrating the diffraction pattern in \textbf{a,b} along the $[1\overline{1}0]$ direction and phase-sorting the electron arrival times with respect to the optical modulation. The electron intensity in each panel is normalized to the maximum value. \textbf{d,e} Vibration amplitude (angular line shift, \textbf{d}) and phase (\textbf{e}) as a function of the drive detuning, showing the stretched resonance and hysteresis of a Duffing oscillator. Green/violet traces denote experimental upward/downward frequency sweeps; dashed lines: Duffing model fit. \label{fig:2}}
\end{figure}

Figure \ref{fig:2}a shows a close-up of the static diffraction signal within the (880) disk, exhibiting the line-like Bragg feature. Optically exciting the membrane with an intensity modulated periodic train of optical pulses (1 mW optical power) at a modulation frequency of 1.05470 MHz (close to a vibrational resonance of the membrane), the temporally averaged Bragg line broadens markedly (Fig. \ref{fig:2}b), which can be understood from the optically induced vibrational motion of the membrane at the electron probing position, rocking back and forth the local membrane tilt.

To resolve this motion in time, we bin the individual electron events into phase intervals over the modulation cycle (binning width of 10 ns) and analyze the phase-resolved line profiles by integrating along the horizontal direction in Fig. \ref{fig:2}b. The resulting Bragg line profiles (along the $[\overline{1}\overline{1}0]$ direction) for different modulation frequencies are shown in Figure \ref{fig:2}c. The phase-dependent profiles demonstrate a sinusoidal, phase-locked motion of the membrane. As the modulation frequency is swept across the resonance, the vibration amplitude continuously increases up to about 1.2 mrad until at about 1.05486 MHz the vibrational amplitude abruptly drops to below 0.01 mrad. The timing within each oscillation cycle at which the maximum amplitude is reached shifts with the modulation frequency, and is indicated by a green tick in each panel of Fig. \ref{fig:2}c. For videos of the Bragg line oscillation at different condition, see Supplemental Video 2. A detailed analysis of the relation between the angular shift of the Bragg lines, the vibrational mode amplitudes, and the mode shapes is given in the Supplementary Information.

Fitting the Bragg line profiles to Lorentzian curves and extracting vibrational amplitude and phase (see Supplementary Information), we obtain the frequency-dependent structural response curve of the membrane, as shown in Fig. \ref{fig:2}d. Notably, the structural response depends on the direction of the frequency sweep with a larger maximum deflection reached for a frequency up-sweep (green curve) as compared to a down-sweep (violet curve). In addition, the asymmetric response curve markedly deviates from a Lorentzian response of a driven harmonic oscillator. Instead, both the stretched response curve and the bistable behavior are well described within a Duffing oscillator model, $\left (\partial_t^2+\omega_0^2 \right) u+\alpha_{\text{Duff}} u^3+2\mu\partial_t u=E(t)$, in which $u$ is the mode amplitude, $\omega_0$ its (angular) resonance frequency in the harmonic limit, $\mu$ a linear friction coefficient and $E(t)=E_0\sin(\omega t)$ a sinusoidal driving force linked to the experimental repetition rate, i.e. $\omega=2\pi/T$. The coefficient $\alpha_{\text{Duff}}$ denotes the effective Duffing coefficient, adding a non-linearity to the restoring force of a simple harmonic oscillator. Since CBED probing is sensitive to the local crystal orientation, the proportionality of the Bragg line shift $\Delta \theta$ to the mode amplitude $u$ depends on the probing location and the mode shape. For the fundamental drumming mode of the central circular membrane studied here and for an electron probing position optimized for maximum tilt amplitude, an approximate connection is given by $\Delta \theta\approx u/R$, with the inner membrane radius $R=50$~µm. Both, the experimental up- and down-sweep data are quantitatively reproduced by a perturbative solution of the Duffing model (dashed black line, see Methods) adopting $\omega_0/(2\pi)=1.05460$~MHz, $\mu/(2\pi)= 4.4 $~Hz, $\alpha_{\text{Duff}}=7.4\,\textrm{kHz}^{2}/\textrm{nm}^2$ and $E_0=22.3 \,\textrm{m}\textrm{s}^{-2}$. 

Further experimental data for lower optical powers and corresponding analyses are given in the Supplementary Information. Notably, due to broken mirror symmetry of the Si/SiO$_x$N$_y$ stack, quadratic and cubic contributions are symmetry allowed for our resonator. As discussed in the Supplementary Information, the lowest order correction for a Duffing nonlinearity does not discriminate between an added quadratic, $\alpha_2u^2$, or cubic, $\alpha_3u^3$ nonlinearity. From the observed DC offset in $u$ at larger driving amplitude, we infer that quadratic and cubic contributions to $\alpha_\text{Duff}$ are similar in magnitude. 
Taking into account the contribution from the quadratic restoring force and the experimentally determined nonlinear coefficient $\alpha_{\text{Duff}}$, we can estimate the cubic contribution of $\alpha_3=17.5$ kHz$^2$/nm$^2$, which is in reasonable agreement with an analytical expression for the nonlinearity in a circular membrane given by $\alpha\approx \frac{Y}{0.27 R^4 \rho}=33$ $\textrm{kHz}^2/\textrm{nm}^2$, with $Y$ and $\rho$ being the average Young's modulus and the mass density of silicon \cite{Davidovikj.2017}. The low damping constant $\mu$ corresponds to a mode quality factor of larger than $10^5$, consistent with ring-down measurements in the low-power limit (see Supplementary Information). 

With the same set of adapted model parameters, the phase relation between the driving optical pulse train and the membrane vibration, as shown in Fig. \ref{fig:2}e, is also quantitatively reproduced. We note that different from a harmonic oscillator, a Duffing oscillator does not achieve a $\pi/2$ phase shift at the resonance frequency $\omega_0$ but instead at the cut-off point in the response curve, at which the vibration amplitude suddenly drops to small values (here at a detuning $\Delta f=\frac{\omega-\omega_0}{2\pi}=240$~Hz). 

\subsection*{Strongly Nonlinear Regime}\label{sec4}

\begin{figure}[!htbp]
    \centering
    \includegraphics[width = 1\textwidth]{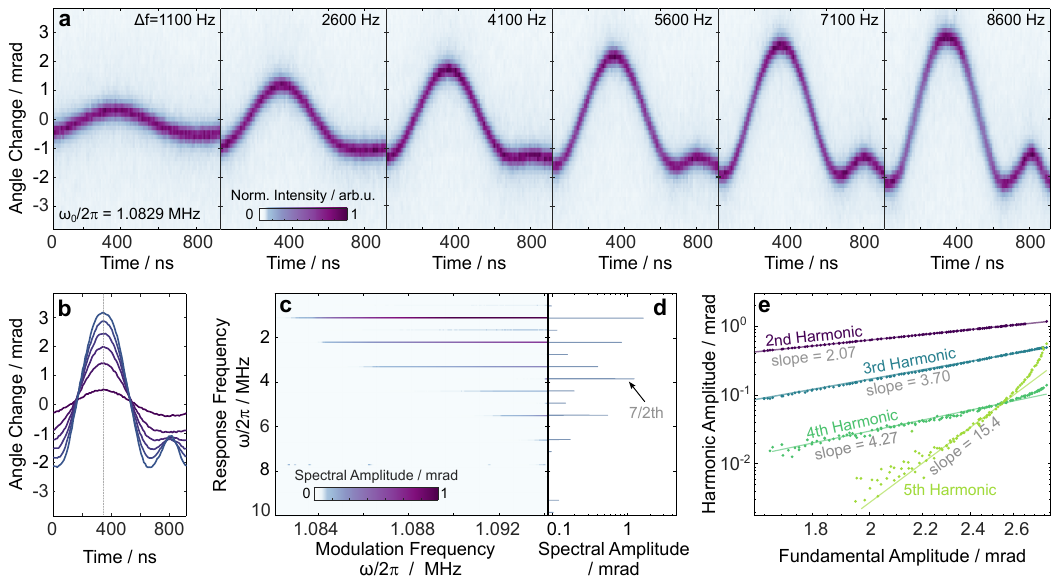}
    \caption{ \textbf{Strongly nonlinear response regime and spectral content at higher drive amplitudes.} \textbf{a} Phase-resolved Bragg line profiles at an excitation power of 17 mW over one drive period for different detunings $\Delta f$ (resonance frequency $\omega_0/(2\pi)=1.082875$ MHz). \textbf{b} Extracted relative Bragg line positions for the traces shown in \textbf{a}. \textbf{c} Fourier transform map of the extracted line profiles for each drive frequency showcasing the emergence of higher harmonics in the structural response. \textbf{d} Frequency spectrum of the Bragg line dynamics at a detuning of 11252~Hz. \textbf{e} Amplitudes of selected higher harmonic components shown in \textbf{c} plotted as a function of the fundamental amplitude. \label{fig:3}}
\end{figure}

Within the Duffing model, both the frequency width and the maximum amplitude of the nonlinear resonance are expected to increase with higher driving strength $E(t)$. As shown in Fig.~\ref{fig:3}a, such a behavior is qualitatively found in the observed membrane bending at an optical excitation power of 17~mW, for which a resonant response occurs over a detuning range of larger than 8.6~kHz and reaches amplitudes of about 3 mrad.  However, similar to indications in previous reports \cite{Jong.2023, Yang.2019, Yang.2021}, the vibrational motion does no longer follow a pure sinusoidal trajectory, which becomes particularly pronounced for the largest amplitudes. For quantifying the change in the temporal trace, we again extracted the phase-dependent Bragg line position by a Lorentzian fit (Fig.~\ref{fig:3}b), and retrieved the individual Fourier components, as depicted in Fig. \ref{fig:3}c,d, depending on the modulation frequency detuning $\Delta f$. For large vibration amplitudes, higher harmonics of the driving frequency are indeed observed up to the 7th order.

The scaling behavior of the higher-harmonic amplitudes with respect to the fundamental amplitude is shown in Fig. \ref{fig:3}e. In a double-logarithmic plot, the amplitudes of the second, third and fourth harmonic increase with a slope of 2.07, 3.70 and 4.27, respectively, indicating an approximately quadratic, cubic and quartic scaling with the fundamental amplitude. In principle, such harmonics could be induced by higher-order terms in the restoring force of a generalized Duffing oscillator. However, as detailed in the Methods, such a nonlinear single-mode model cannot fully describe the experimental data since the simultaneously expected rectification component (resulting in a static line shift) is largely suppressed. In particular, while the cycle-averaged line shift scales quadratically with the fundamental vibration amplitude, it remains below 80~µrad at the strongest drive, corresponding to a membrane displacement of only 4 nm, still above the few-\AA~noise floor of the measurement. This upper bound would imply a second harmonic amplitude below 27~µrad, contrary to the experimentally observed 2$\omega$-amplitudes of up to 1~mrad. We note that different from the often employed probing by laser Doppler vibrometry \cite{Rembe.2007}, time-resolved CBED is sensitive to static membrane displacements. 

Instead, we can quantitatively describe the evolution of the bending trajectories across the resonance, by considering a nonlinear multi-mode model using the same parameters $\alpha_{\text{Duff}}$ and $\mu$ extracted in the low-power regime. In this model the driving force $E(t)$ is scaled linear with the applied optical power and nonlinear couplings to vibrational modes at higher frequencies $\omega_{0,i}$ with amplitudes $u_i$ are included through mode coupling potentials of the form $V=\sum_{i\geq 2}\lambda_i u_1^i u_i$. If, for example, the first overtone of the driving frequency, $2\omega$, lies slightly below the resonance frequency $\omega_{0,2}$ of a higher order mode, the second harmonic contribution to the membrane bending is in phase with $u$ and can be resonantly enhanced without requiring a correspondingly large DC component. In addition, for cases sufficiently far from resonance, the scaling behavior of the higher-harmonic amplitudes is approximately given by $u_n \propto u_1^n $, in agreement with the experimental data.

Whereas at the chosen probing position the line shift proceeds predominantly along the $[\overline{1}\overline{1}0]$ direction, we also start to observe additional line movement in the orthogonal $[1\overline{1}0]$ direction, with primary frequency components at $4\omega$ and $5\omega$, as discussed in the Supplementary Material.  

\subsection*{Period Doubling and Complex Phase Trajectories}\label{sec5}
Interestingly, for transients with the highest membrane bending amplitude, the local structural response exhibits not only integer but also half-integer spectral components, most prominent at the $7/2$-th harmonic at about 3.83 MHz  (arrow Fig. \ref{fig:3}d, right panel). The appearance of such additional Fourier components highlights that the vibrational dynamics does no longer follow the periodicity of the driving field.

\begin{figure}[!htbp]
    \centering
    \includegraphics[width = 1\textwidth]{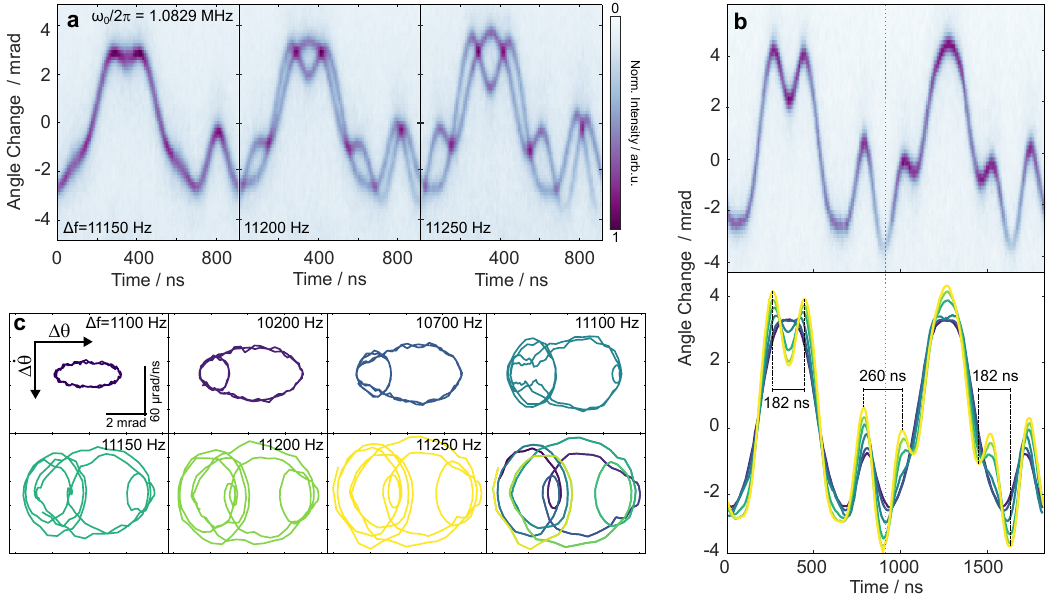}
    \caption{ \textbf{Period-doubling bifurcation and complex phase-space trajectories.} \textbf{a} Phase-resolved Bragg line profiles at large detunings and high driving amplitude (17 mW), highlighting the Bragg line splitting due to a period doubling bifurcation in the membrane response. \textbf{b} Top panel: Bragg line profile from the rightmost panel in \textbf{a}, unfolded into a time interval twice of the driving period. Bottom panel: Extracted period-doubled, relative Bragg line positions for different detunings (10200 - 11250 Hz, violet to yellow, same color coding as the corresponding trajectories in \textbf{c}). Gray dashed line indicates one period of the external drive. Indicated time intervals of 182 and 260 ns align with the high-amplitude Fourier components for the 5th and 7/2th harmonics, respectively. \textbf{c} Trajectory maps in ($\Delta\theta,\Delta\dot{\theta}$)-space for different drive detunings. For better visualization of the dynamic behavior at high frequencies, the bottom right panel is color coded to indicate the temporal evolution from violet to yellow.\label{fig:4}}
\end{figure}

This effect becomes fully visible at further increased vibrational amplitudes, as shown in Fig.~\ref{fig:4}a. Here, the phase-resolved Bragg line profile exhibits a pronounced line splitting, increasing with the vibrational amplitude. Since our time-resolved imaging methodology involves phase-resolved averaging over many oscillation cycles in which the electron event stream is back-folded onto a single driving period $T$, such a line splitting could in principle be the result of an incoherent switching between two different states of the driven system within the acquisition time. However, by applying a back-folding to twice the driving period (see Methods), the line splitting completely vanishes, as shown in Fig.~\ref{fig:4}b (top panel), ruling out the possibility of an incoherent mixture of states and demonstrating the presence of a stable period doubling dynamics. Representative phase-resolved Bragg-line dynamics are shown in the Supplementary Movie 2.

The observed period-doubling represents a spontaneous breaking of the temporal symmetry of the $T$-periodic equation-of-motion (EOM), so that in principle transients $u(t)$ and $u(t+T)$ are solution of the EOM. Remarkably, the period-doubled response stably follow one of these solutions over the full acquisition time of 2~s, corresponding to about $2\times 10^6$ oscillation periods. As seen in the extracted line position (Fig. \ref{fig:4}b, bottom panel), the prominent modulation period appearing during period doubling corresponds to about 260~ns, in line with the largest non-integer harmonic in Fig. \ref{fig:3}d. 

In contrast, a conventional parametric modulation would result in a dominant frequency component at $\omega/2$. As discussed in the Supplementary Information, different nonlinear mode couplings, such as $\propto u_1^7u_{7/2}^2$, would lead to a parametric driving of a mode with a resonant frequency at about $(7/2)\omega$. 

Finally, it is instructive to represent the experimental Bragg line trajectories along $[1\overline{1}0]$ in $(\Delta\theta,\Delta\dot{\theta})$-space, as displayed in Fig.~\ref{fig:4}c. At small oscillation amplitudes (Fig.~\ref{fig:4}c, upper leftmost panel), the movement follows the well-known elliptical phase-space trajectory of an harmonic oscillator. The higher-harmonic components at intermediate amplitudes are visible in the back-folding of one end of the ellipse (upper middle panel). For further increased amplitude, the elliptical trajectory first develops a pronounced kink (upper rightmost panel) from which the doubling bifurcation develops.

\subsection*{Conclusion}\label{sec6}
In conclusion, we demonstrated the capabilities of ultrafast transmission electron microscopy for locally probing the nonlinear dynamics of a resonantly driven prototypical nanomechanical oscillator. In an optically driven silicon membrane, we resolved the evolution from Duffing bistability to higher-harmonic generation and period doubled dynamics. The diffraction-based methodology combines nanosecond temporal resolution with local probing on the nanometer scale and resolves displacement amplitudes down to a few \AA\ through its response to local membrane bending and displacement gradients. It is equally applicable to single-crystal resonators down to length scales of a few nanometers and can  be also adapted for electrical or mechanical driving schemes. As such, we believe that this approach can help to push nonlinear nanomechanics into the atomic-scale regime. A particular strength of the diffractive probing is its sensitivity to structural displacement gradients which remain sizable even for nanoscale oscillators. Finally, we envision that the large structural distortion observed in our membrane device could be beneficial as a driver for phase transition in materials deposited on the membrane, opening up the investigation of strain-driven ultrafast nanoscale dynamics. 

\section*{Methods}

\subsection*{Event-based convergent beam electron diffraction and data analysis}
The experiments were conducted using the Regensburg UTEM, which is based on a JEOL JEM F200 TEM, operated at an acceleration voltage of 200 kV. The microscope is equipped with a laser driven cold field emitter gun \cite{Schroder.2025}, but all measurements reported here were carried out in continuous field-emission mode. Convergent-beam electron diffraction (CBED) was recorded at a nominal camera length setting of 2000 mm. The electron probe was focused down to a spot size of approximately 20 nm (FWHM) at the sample with a semi-convergence angle of 5 mrad, and aligned at an angle of 1.5 degree relative to the Si [001] zone axis.

The silicon membrane was optically excited using a continuous-wave diode laser (785 nm, iBeam Smart, Toptica) which was digitally amplitude-modulated synchronous to the output of a frequency generator (33622A, Keysight). The modulation frequency is tunable up to 250 MHz with sub-Hz precision at an amplitude rise time of 2.9 ns. The laser beam was focused onto the membrane using a lens outside of the TEM vacuum, resulting in a focal spot diameter of approximately 20 µm. The optical power at the sample surface was varied between 1 and 17 mW.

Diffraction patterns were recorded using a pixelated electron detector (CheeTah T3, Amsterdam Scientific Instruments, 512 x 512 pixels, sensor size of 28 mm x 28 mm), mounted close to the projection lens. This event-based detector is based on the Timepix3 architecture, enabling single-electron sensitivity and time-stamps each detected electron with nanosecond precision. In addition, a reference trigger from the function generator that modulates the laser amplitude was fed into the Timepix3 timing input and recorded as additional digital events. Using this timing reference, individual electron arrival times were phase-sorted relative to the optical excitation. Post-processing of the event stream yielded phase-resolved CBED frames across the modulation cycle.

For data analysis, the event stream was binned into 100 phase intervals per drive cycle using an electrical reference signal synchronous with the laser modulation. For each phase bin, CBED images were reconstructed by accumulating all events per pixel. Bragg-line profiles were obtained by integrating within a region of interest perpendicular to the line direction. Profiles were fitted with Lorentzian functions to extract the line center, width, and amplitude for each phase $\phi$. The angular displacement $\Delta \theta$ was calculated from the calibrated pixel-to-angle conversion factor of 6.25 $\times 10^{-5} \textrm{rad}~\textrm{px}^{-1}$. Frequency sweeps were performed in 1-Hz steps with a dwell time of 1--2 s, and both upward and downward sweeps were recorded to capture the resonance hysteresis.

To analyze the emergence of spectral components at high driving strength, the phase-sorted Bragg-line position traces were transformed into the frequency domain. For each driving frequency $\omega$, the measured radial motion $\theta(t)$, evaluated over the interval $t=[0,4\pi/\omega[$ was Fourier transformed after zero-padding to improve visualization of the spectrum. From each spectrum, we extracted the Fourier amplitudes at the fundamental response frequency $\omega_0$ and at selected higher-order and fractional multiples of the drive frequency. In particular, we tracked the evolution of the components $\omega$, $2\omega$, $3\omega$, $4\omega$, $5\omega$ , $7\omega$ and the half-integer $7/2 \omega$ component, as the frequency was swept across the resonance. 

\subsection*{Theoretical Modeling -- Duffing Oscillator}
For small optical driving power, we describe the structural dynamics of the silicon membrane vibration within a simple Duffing oscillator model for which the time-dependent amplitude of a vibrational mode is given by  

\begin{equation}
\left (\partial_t^2+\omega_0^2 \right) u+\alpha u^3+2\mu\partial_t u=E_0 \sin \omega t.
\label{eq:DuffingEquation}
\end{equation}

For small amplitudes, the dynamics can be analyzed within the harmonic balance approach \cite{Nayfeh.1979,Kosata.2022}, which, in the lowest order, considers an ansatz for the amplitude in the form $u=a(t) \sin \omega t + b(t) \cos \omega t$, arriving at a set of nonlinear differential equations in $a$ and $b$. Neglecting second-order derivatives and evaluating stationary solutions of the Fourier components of $a$ and $b$ at frequencies $\omega$, one obtains an implicit analytical solution given by
\begin{align}
\begin{split}
E_0 &=a \left (\omega_0^2-\omega^2 \right) + 2 b \mu \omega +\frac{3}{4}\alpha\left ( a^3+b^2 a\right )\\
0 &=b \left (\omega_0^2-\omega^2 \right) -2 a \mu \omega +\frac{3}{4}\alpha\left ( b^3+a^2 b\right ) 
\label{eq:DuffingImplicit}
\end{split}
\end{align}
and the vibrational amplitude $u_0=\sqrt{a^2+b^2}$ can be obtained from
\begin{equation}
    E^2=u_0^2 \left[(\omega_0^2-\omega^2+\frac{3}{4}\alpha u_0^2)^2+(2\mu\omega)^2  \right].
    \label{eq:DuffingFit}
\end{equation}

Driving the system close to resonance, i.e. at a small detuning $\sigma=2\pi \Delta f=\omega-\omega_0\ll \omega_0,\omega$, and small anharmonicity, Eq.~\ref{eq:DuffingFit} can be further simplified, yielding
\begin{equation}
\sigma=\frac{3}{8}\frac{\alpha u_0^2}{\omega_0} \pm \sqrt{\frac{E_0^2}{4\omega_0^2 u_0^2}-\mu^2}.
\label{eq:duffingAmplitude}
\end{equation}

Thus, for a detuning sweep, the maximum vibration amplitude is obtained when the root in Eq.~\ref{eq:duffingAmplitude} vanishes, i.e. for $u_{0,\text{max}}=\frac{E_0}{2\omega_0 \mu}$, and,  for varying driving strengths, the detuning at which the maximum amplitude is found traces the curve $\sigma=\frac{3}{8}\frac{\alpha}{\omega_0}u_{0,\text{max}}^2$. 

To extract the Duffing parameters from the measured frequency sweeps (Figs. 2d,e and S4), we first determine the amplitude $u_0(\omega)$ by fitting the experimentally obtained oscillation $u(t)$ with a sinusoidal model $u(t)=u_0\sin(\omega t-\phi)$. The Duffing parameters $\omega_0,\alpha,\mu$ and $E_0$ are obtained from a nonlinear least-squares solver, minimizing the normalized residuals between the experimental frequency-dependent amplitudes $u_0(\omega)$ and the implicit Duffing expression Eq.~\ref{eq:DuffingFit}:
\begin{equation}
    r=\frac{u_0^2[(\omega_0^2-\omega^2+3/4\alpha u_0^2)^2+(2\mu\omega)^2]-E_0^2}{E_0^2}
\end{equation} 

Whereas, so far, we only considered a cubic nonlinearity in the restoring force, $\alpha u^3$, Eq.~\ref{eq:DuffingFit} is equally fulfilled (in the lowest order harmonic balance approach) if an additional quadratic restoring force, $\alpha_2 u^2$, is present. In this case, $\alpha$ is replaced by $\alpha_{\text{Duff}}=\alpha_3-\frac{10 \alpha_2^2}{9 \omega_0^2}$. 

Analogous to optical rectification, such a quadratic term in the restoring force not only induces second harmonic frequency components but also a non-zero offset in the vibration $u(t)$, which is given by
\begin{equation}
    u_\text{offset}=-\frac{\alpha_2 u_0^2}{2\omega_0^2}.
\end{equation}
As shown in the Supplement, we experimentally observe such a negative period-averaged sample tilt that scales quadratically with the oscillation amplitude, $\Delta\theta_{\mathrm{offset}}\propto -\Delta\theta_0^2$, consistent with an additional quadratic term in the restoring force. Using the maximum observed offset, $\Delta\theta_{\text{offset}}=80$~µrad at $\Delta \theta=$2.7 mrad, we estimate $\alpha_2 \approx 2.03 \times 10^{4}$ kHz$^{2}$/nm. Thereby, the coefficient of the cubic contribution is on the order of $\alpha_3=\alpha_{\text{Duff}}+10\alpha_2^2/(9\omega^2)\approx 17.5 $ kHz$^{2}$/nm$^2$, i.e. about twice as large as the nonlinear coefficient extracted from the Duffing fit. Therefore, the measured nonlinear response is consistent with a combination of quadratic and cubic restoring-force nonlinearities. We stress that the $\alpha_2$ extracted here accounts only for the weak rectification offset; the large second-harmonic amplitude observed experimentally is not captured by this single-mode term and is instead attributed to resonant multimode coupling (see Results and SI Note 8).

\section*{Data availability}
The data that support the findings of this study are available from the corresponding author upon reasonable request.



\section*{Acknowledgments}
We acknowledge financial support from the Volkswagen Foundation as part of the Lichtenberg Professorship ‘‘Ultrafast nanoscale dynamics probed by time-resolved electron imaging’’. Funding by the Deutsche Forschungsgemeinschaft (DFG, German Research Foundation) through INST 184/211-1 FUGG, GRK 2905 (project-ID 502572516) and CRC 1277 (project-ID 314695032, subproject C02) is gratefully acknowledged. Further partial funding is provided by the Free State of Bavaria through the Lighthouse project ‘‘Free-electron states as ultrafast probes for qubit dynamics in solid-state platforms’’ within the Munich Quantum Valley initiative.

\section*{Author contributions}
A.S. and K.N. carried out TEM experiments and data analysis. A.S. and S.S. initiated the project and wrote the manuscript. All authors contributed to the interpretation of the data and the writing of the manuscript.

\section*{Competing interests}
The authors declare no competing interests.

\end{document}